\documentclass[onecolumn,showpacs,aps,floatfix,pra,superscriptaddress,nofootinbib]{revtex4}
\usepackage{color}
\usepackage{graphicx}
\usepackage{bm}
\usepackage{amsmath}
\usepackage{enumerate}
\usepackage{upgreek}
\newcommand{\tr}{ \text{tr} }
\newcommand{\re}{ \text{ref} }
\newcommand{\erf}{ \text{erf} }
\newcommand{\erfc}{ \text{erfc} }
\newcommand{\Ima}{ \text{Im} }

\newcommand{\CK}{ \text{CK} }
\newcommand{\xd}{ \text{d} }
\newcommand{\ti}{ \tilde }

\newcommand{\pa}{ \partial }
\newcommand{\hb}{ \hbar }
\newcommand{\si}{ \sigma }
\newcommand{\ga}{ \gamma }
\newcommand{\om}{ \omega }
\newcommand{\Om}{ \Omega }
\newcommand{\la}{ \langle }
\newcommand{\ra}{ \rangle }
\begin{document}
\title{Characteristic times for thermal wave packets in dissipative Bohmian mechanics: the parabolic repeller}
\author{S. V. Mousavi}
\email{vmousavi@qom.ac.ir}
\affiliation{Department of Physics, University of Qom, Ghadir Blvd., Qom 371614-6611, Iran}
\author{S. Miret-Art\'es}
\email{s.miret@iff.csic.es}
\affiliation{Instituto de F\'isica Fundamental, Consejo Superior de
Investigaciones Cient\'ificas, Serrano 123, 28006 Madrid, Spain}
\begin{abstract}
Thermal wave packets are used to analyze transmission probabilities and characteristic times through a parabolic repeller
within the dissipative Bohmian mechanics. Thermal arrival, dwelling, transmission and reflection times are defined and calculated by 
using a Maxwell-Boltzmann distribution for the initial velocities of the incident particles. The dissipation is considered within the 
Caldirola-Kanai and Kostin approaches where a linear and nonlinear framework is used, respectively. The initial parameters are chosen to 
have only a dissipative tunnelling dynamics at zero temperature; at any nonzero temperature, transmission proceeds not only via tunnelling.

\end{abstract}
\maketitle


%========================================
\section{Introduction}
%========================================

Time in quantum mechanics is a permanent and important issue subject to many definitions depending on the process studied, 
leading to very interesting and endless debates for conservative problems \cite{Landauer,Muga-book1, Muga-book2}. This is due mainly to 
the fact that time is usually considered as a parameter (external parameter)  and not as an observable in the non-relativistic framework. Thus, we
usually talk about phase, tunnelling, transmission, resident, dwelling, arrival, etc. times. In general, we could globally named them as characteristic
times. Calculations of these characterisitic times are usually extracted from some time distributions. For example, the so-called transition path time
distribution which gives the probability distribution of transition times between two spatial points has been proposed in the context of the transition
state theory and rate coefficients \cite{Hummer}. This distribution defined in terms of a symmetrized thermal density correlation function 
has been succesfully used by Pollak to calculate tunnelling times in presence of (Ohmic) friction following the Caldeira-Leggett Hamiltonian  
in the Langevin formalism \cite{Eli1,Eli2,Po-PRL-2017,Po-PRA-2017}.

On the other hand, Ford, Lewis and O'Connell \cite{FoLeCo-PLA-1988,FoLeCo-PRA-2001,FoCo-PLA-2001,FoCo-AJP-2002,FoCo-JOB-2003} 
studied decoherence in the quantum {\it Brownian} motion at high temperatures in the absence of dissipation and a zero temperature with 
dissipation starting from the quantum Langevin equation for Ohmic friction and for thermal wave packets.  In particular, in Ref. 
\cite{FoLeCo-PLA-1988} they studied tunnelling through a parabolic potential in the presence of dissipation. 

Within the framework of Bohmian mechanics \cite{Holland-book-1993}, where particle trajectories are calculated, time quantities 
like arrival time, transmission and reflection times are unambiguously defined and widely used for conservative dynamics 
\cite{Le-book-2002,Le-LNP-2008}. 
In continuation of our study of dissipative tunnelling through a parabolic repeller \cite{MoMi-JPC-2018,MoMi-AP-2018} through scaled trajectories, 
we extend this study here by introducing temperature through the Maxwell-Boltzmann distribution of velocities for an incoming wave packet 
and analyzing some typical characterisitic times. Our system represents thus an ensemble of non-interacting particles with a thermal distribution 
of initial velocities. The initial state is taken to be a mixed ensemble of Gaussian wave packets with weights given by the Maxwell-Boltzmann
distribution as Ford et al. \cite{FoCo-AJP-2002} and also in the study of time-of-flight distribution for a cloud of cold atoms falling freely 
under gravity \cite{MaHoMaPa-PRA-2007}. Thermal arrival, dwelling, transmission and reflection times are defined and calculated 
through dissipative Bohmian trajectories. Dissipation is considered within the Caldirola-Kanai and Schr\"odinger-Langevin or Kostin approaches 
\cite{NaMi-book-2017}, within a linear and nonlinear theoretical framework, respectively.
Leavens \cite{Le-PLA-1993} has also shown that arrival time distributions are given by the modulus of the probability 
current density. Due to the non-crossing property of Bohmian trajectories, in a scattering (transmission) problem there is a critical 
trajectory which bifurcates transmitted trajectories from the reflected ones. This provides a way to split dwelling time into transmission and 
reflections times \cite{Le-LNP-2008}. In this way, the computation of characteristic times reduces to the computation of the single critical 
trajectory. However, it is also proved \cite{Kr-JPA-2005} that there is even no need to compute this single trajectory.

The paper is organized as follows. In Section  \ref{sec: thermal_wave}, thermal wave packets are built for an ensemble of non-interacting 
particles, each one being described by a Gaussian wave packet whose central velocity is distributed according to the Maxwell-Boltzmann 
distribution function. In section \ref{sec: dissipation}, the effect of the dissipation on the evolution of the thermal wave packet is considered
within the Bohmian mechanics framework. Finally, Section \ref{sec: tran_rep} presents and discusses transmission through a parabolic repeller 
by considering different characteristic times such arrival, dwelling and reflection times.

%=====================================================
\section{ Constructing thermal wave packets } \label{sec: thermal_wave}
%=====================================================

Consider an ensemble of noninteracting particles where each particle is initially described by the state $ | \psi_{v_0}(0) \rangle $ 
with $v_0$ being the central velocity of the corresponding wave packet 
\begin{eqnarray} \label{eq: psi_gauss_0}
\psi_{v_0}(x, 0) &=& \frac{1}{(2\pi \si_0^2)^{1/4}} \exp \left[ -\frac{(x-x_0)^2}{4 \si_0^2} + i \frac{m v_0}{\hb} x \right] .
\end{eqnarray}
Particles are assumed to have a Maxwell-Boltzmann distribution of initial velocities given by
\begin{eqnarray} \label{eq: Maxwell_Boltz_dis}
f_T(v_0) &=& \sqrt{ \frac{m}{2\pi k_B T} } \exp \left[ -\frac{m v_0^2}{2 k_B T} \right]    ,
\end{eqnarray}
$m$ being the mass of the particles, $k_B$ the Boltzman constant and $T$ the temperature.

According to Eq. (\ref{eq: mixed_state_0}) of Appendix \ref{app: den_mat}, our mixed ensemble is described by \cite{FoCo-AJP-2002}
\begin{eqnarray} \label{eq: den_mat_0}
\hat{\rho}_T(0) &=& \int_{-\infty}^{\infty} dv_0 ~ f_T(v_0) | \psi_{v_0}(0) \rangle \langle \psi_{v_0}(0) | 
\end{eqnarray}
whose time evolution is given by the von-Newmann equation of motion (\ref{eq: voNewmann}) and expressed as
\begin{eqnarray} \label{eq: den_mat_t}
\hat{\rho}_T(t) &=& e^{-i \hat{H} t / \hb } \hat{\rho}_T(0) e^{i \hat{H} t / \hb } =
\int_{-\infty}^{\infty} dv_0 ~ f_T(v_0) ~ | \psi_{v_0}(t) \rangle \langle \psi_{v_0}(t) |
\end{eqnarray} 
where $| \psi_{v_0}(t) \rangle = e^{-i \hat{H} t / \hb } | \psi_{v_0}(0) \rangle$, $H$ being the Hamiltonian of the system.

%
%In order to avoid dissipation in the equation of motion,  we have supposed the ensemble is weakly coupled to a heat bath \cite{FoCo-AJP-2002}.
%
Matrix elements of the thermal density operator (\ref{eq: den_mat_t}) in the coordinate representation are given by 
\begin{eqnarray} \label{eq: mat_elem_t}
\rho_T(x, x', t) &=&  
\int_{-\infty}^{\infty} dv_0 ~ f_T(v_0) ~ \psi_{v_0}(x ,t) \psi^*_{v_0}(x', t) .
\end{eqnarray}

In the following, we will separately consider propagation in the force-free field, propagation in a constant force field and in a linear force field 
where the propagators are known to have analytic expressions. We find that, in all cases considered, the diagonal elements of the density matrix,
which are interpreted as a probability distribution, has the Gaussian form
\begin{eqnarray} \label{eq: diag_mat_elem_t}
\rho_T(x, t) &=&  \rho_T(x, x', t) \bigg|_{x'=x}
= \frac{1}{\sqrt{2\pi} \si_T(t)} \exp\left[{-\frac{( x - X(t) )^2}{2 \si_T(t)^2}} \right]
\end{eqnarray}
where the center of the packet follows the thermal-averaged trajectory $ X(t) = \la x_t \ra $, $x_t$ being the center of the wave packet $ |\psi_{v_0}(x, 0)|^2 $; and its width has a temperature contribution.
From (\ref{eq: obser_expect}) ,we have that 
\begin{eqnarray} \label{eq: thermal_expect_value}
\langle \hat{A} \rangle_T(t) &=&  \int_{-\infty}^{\infty} dv_0 ~ f_T(v_0) 
\int_{-\infty}^{\infty} dx ~ \psi^*_{v_0}(x, t) ~ A\left(x, -i\hb \frac{\pa}{\pa x} \right) ~  \psi_{v_0}(x ,t)
\end{eqnarray}
for the expectation value of an observable $\hat{A} = \hat{A}(\hat{x}, \hat{p})$.
As a special case, the expectation value of space coordinate $\hat{x}$ is given by
\begin{eqnarray} \label{eq: thermal_x_expect_value}
\langle \hat{x} \rangle_T(t) &=&  
\int_{-\infty}^{\infty} dv_0 ~ f_T(v_0) \int_{-\infty}^{\infty} dx ~ \psi^*_{v_0}(x, t) ~ x ~  \psi_{v_0}(x ,t) \nonumber \\
&=&
\int_{-\infty}^{\infty} dx ~ x ~ \rho_T(x, t) 
\end{eqnarray}
This relation confirms the interpretation of the diagonal matrix elements of the density operator as the probability density.

%=====================================================
\subsection{Free particles}
%=====================================================
The propagator of the free particle, $\hat{H} = \frac{\hat{p}^2}{2m}$, is given by
\begin{eqnarray} \label{eq: free_prop}
\langle x | e^{-i \hat{H} t / \hb } | x' \rangle &=&  
\sqrt{ \frac{m}{2\pi i \hb t} } \exp \left[ \frac{i m}{ 2 \hb t} (x-x')^2 \right] 
\end{eqnarray}
and the wave function with initial velocity $v_0$ is then
\begin{eqnarray} \label{eq: wave_t_free}
\psi_{v_0}(x, t) &=& \int_{-\infty}^{\infty} dx' ~ \langle x | e^{-i \hat{H} t / \hb } | x' \rangle ~
\psi_{v_0}(x', t)  
\nonumber \\
&=& \frac{1}{(2\pi)^{1/4} \sqrt{s_t}}
\exp \left[ \frac{i m}{2 \hb t} \left( x^2 + \frac{i \hbar t}{2 m \si_0^2} x_0^2 \right) - \frac{\si_0}{s_t} \left( x-x_0-v_0 t + \frac{s_t}{\si_0} x_0 \right)^2 \right]
\end{eqnarray}
where the complex width is
\begin{eqnarray} \label{eq: st}
s_t &=& \si_0 \left( 1 + i \frac{\hb t}{2 m \si_0^2} \right) .
\end{eqnarray}
From (\ref{eq: wave_t_free}) ,the Gaussian shape (\ref{eq: diag_mat_elem_t}) for the diagonal elements of density matrix with
\begin{eqnarray} 
X(t) &=& x_0 \label{eq: xt_free} \\
\si_T(t) &=& \si_0 \sqrt{ 1 + \left( \frac{ \hb^2 }{4 m^2 \si_0^4} + \frac{k_B T}{m\si_0^2} \right)t^2 } 
\label{eq: sigmat_free}
\end{eqnarray}
is obtained.

As one clearly sees there is a temperature-dependence contribution to the width. For a given time, the width increases with temperature. 
Now, from Eq. (\ref{eq: thermal_expect_value}), the first two momenta  of the momentum distribution for a given temperature are
\begin{eqnarray} 
\langle \hat{p} \rangle_T(t) &=& 0 \label{eq: p_ex_free} \\
\langle \hat{p}^2 \rangle_T(t) &=& \frac{\hb^2}{4\si_0^2} + m k_B T \label{eq: p2_ex_free}
\end{eqnarray}
and the uncertainty is then given by
\begin{eqnarray} \label{eq: p_uncer_free}
\Sigma_T &=& \sqrt{ \langle \hat{p}^2 \rangle_T(t) - \langle \hat{p} \rangle^2_T(t) } = \sqrt{ \frac{\hb^2}{4\si_0^2} + m k_B T } 
\end{eqnarray}
which is time-independent but has a temperature-dependence contribution. By using (\ref{eq: j}) or (\ref{eq: cur_mixed}), the thermal probability current density can be expressed as
\begin{eqnarray} \label{eq: cur_para}
j_T(x, t) &=& \left[ 
\left( \frac{\Sigma_T}{m ~\si_T(t)}  \right)^2 (x-x_0) ~ t \right]   .
~\rho_T(x, t)
\end{eqnarray}
As a consistency check, the thermal Wigner distribution function which is defined by
\begin{eqnarray} \label{eq: wigner}
W_T(x, p, t) &=& \frac{1}{\pi \hb } \int_{-\infty}^{\infty} dy~ 
\langle x+y | \hat{\rho}_T(t) | x-y \rangle
e^{i 2 p y/ \hb } \nonumber \\
&=&
\frac{1}{\pi \hb } \int_{-\infty}^{\infty} dv_0 ~ f_T(v_0) \int_{-\infty}^{\infty} dy ~
\psi_{v_0}(x+y, t) ~ \psi^*_{v_0}(x-y, t) e^{i 2 p y/ \hb } 
\end{eqnarray}
can be calculated for the free case reaching
\begin{eqnarray} \label{eq: W_free}
W_T(x, p, t) &=& \frac{1}{ \sqrt{ \pi(\hb^2 + 4 m \si_0^2 k_B T) } } 
\exp \left[ - \frac{2 \si_0^2 p^2}{ \hb^2 + 4 m \si_0^2 k_B T } - 
\frac{ ( m(x-x_0) + p t )^2 }{ 2m^2 \si_0^2 }
 \right].
\end{eqnarray}
By integrating over the spatial coordinate $x$, it leads to the momentum distribution
\begin{eqnarray} \label{eq: mom_dis_free}
\Pi_T(p) &=& \frac{1}{ \sqrt{2\pi} \Sigma_T } \exp \left[ - \frac{ p^2 }{ 2 \Sigma_T^2 } \right]
\end{eqnarray}
with $\Sigma_T$ given by Eq. (\ref{eq: p_uncer_free}).
%=====================================================
\subsection{Linear potential}
%=====================================================
The propagator for the linear potential $V(x) =  K x $ is
\begin{eqnarray} \label{eq: lin_prop}
\langle x | e^{-i \hat{H} t / \hb } | x' \rangle &=&  
\sqrt{ \frac{m}{2\pi i \hb t} } \exp \left[ \frac{i m}{ 2 \hb t} (x-x')^2 - i \frac{K t}{2\hb}(x+x')
- i \frac{K^2}{24 m \hb} t^3 \right] 
\end{eqnarray}
From this, one obtains the Gaussian function (\ref{eq: diag_mat_elem_t}) with a width given by (\ref{eq: sigmat_free}) and the center of the thermal packet follows the trajectory

\begin{eqnarray} \label{eq: xt_linear}
X(t) &=& x_0 - \frac{K t^2}{2 m}    .
\end{eqnarray}
Again, the first two moments of the momentum distribution are 
\begin{eqnarray} 
\langle \hat{p} \rangle_T(t) &=& - K t \label{eq: p_ex_linear} \\
\langle \hat{p}^2 \rangle_T(t) &=& \frac{\hb^2}{4\si_0^2} + m k_B T + K^2 t^2 \label{eq: p2_ex_linear}
\end{eqnarray}
and, thus, the corresponding uncertainty takes again the form of Eq. (\ref{eq: p_uncer_free}).

%=====================================================
\subsection{Parabolic repeller potential}
%=====================================================

The propagator for the inverted parabolic potential $V(x) =  -\frac{1}{2} m \om^2 x^2 $ is
\begin{eqnarray} \label{eq: para_prop}
\langle x | e^{-i \hat{H} t / \hb } | x' \rangle &=&  
\sqrt{ \frac{ m \om }{2\pi i \hb \sinh \om t} } \exp \left[ \frac{i m \om}{ 2 \hb \sinh \om t} ( (x-x')^2 \cosh \om t - 2 x x' ) \right] 
\end{eqnarray}
and the center of the  thermal Gaussian function (\ref{eq: diag_mat_elem_t}) ans its width are in this case
\begin{eqnarray} 
X(t) &=& x_0 \cosh( \om t ) \label{eq: xt_para} \\
\si_T(t) &=& \si_0 \sqrt{
\cosh^2( \om t ) + \left(  \frac{ \hb^2 }{ 4 m^2 \si_0^4 } + \frac{k_B T}{m \si_0^2}\right) \frac{ \sinh^2( \om t ) }{ \om^2 }  .
}\label{eq: sigmat_para}
\end{eqnarray}
Again, from the first two moments of the momentum distribution
\begin{eqnarray} 
\langle \hat{p} \rangle_T(t) &=& m \om x_0 \sinh( \om t )  \label{eq: p_ex_para} \\
\langle \hat{p}^2 \rangle_T(t) &=& \left( \frac{\hb^2}{4\si_0^2} + m k_B T \right) \cosh^2( \om t ) 
+ m^2 (x_0^2 + \si_0^2) \om^2 \sinh^2( \om t )
\label{eq: p2_ex_para}
\end{eqnarray}
the uncertainty is written as
\begin{eqnarray} \label{eq: p_uncer_para}
\Sigma_T(t) &=& \sqrt{ \left( \frac{\hb^2}{4\si_0^2} + m k_B T \right) \cosh^2( \om t ) 
+ m^2 \si_0^2 \om^2 \sinh^2( \om t )
} 
\end{eqnarray}
which is time-independent but  temperature dependent. Finally, the thermal probability current density is
\begin{eqnarray} \label{eq: cur_para}
j_T(x, t) &=& \om \sinh(\om t) ~
\frac{ x\left( \frac{\hb^2}{4\si_0^2} + m k_B T + m^2 \om^2 \si_0^2 \right) \cosh(\om t) - x_0 \left(\frac{\hb^2}{4\si_0^2} + m k_B T \right) }
%{ m^2 ~\si_T(t)^2 }
{ m^2 \si_0^2 \cosh^2( \om t ) + \left(  \frac{ \hb^2 }{ 4 \si_0^2 } + m k_B T \right) \sinh^2( \om t ) }
~\rho_T(x, t)
\end{eqnarray}
%

%===============================================
\section{Dissipation in Bohmian mechanics} \label{sec: dissipation}
%===============================================

We are going to consider dissipation through two different approaches, the linear Schr\"odinger equation coming from the so-called 
Caldirola-Kanai (CK) Hamiltonian  \cite{caldirola, kanai} and the nonlinear, logarithmic Schr\"odinger-Langevin (or Kostin) equation
\cite{Kostin-1972, NaMi-book-2017}, both of them without noise. These equations have been used in our previous works
\cite{MoMi-JPC-2018,MoMi-AP-2018}.

\subsection{Wave equation in the CK and Kostin approaches}

In our context, the Schr\"odinger equation within the CK approach  reads as
\begin{eqnarray} \label{eq: Sch_CK}
i \hbar \frac{\partial}{\partial t}\psi_{v_0, \ga}(x, t) &=& \left[ -\frac{\hbar^2}{2m} e^{-\ga t}
\frac{\partial^2}{\partial x^2} + e^{\ga t}  V(x)   ,
\right] \psi_{v_0, \ga}(x, t) .
\end{eqnarray}
being $\ga$ the friction. For the quadratic potential
\begin{eqnarray} \label{eq: quad_pot}
V(x) &=& K x - \frac{1}{2} m \om^2 x^2   ,
\end{eqnarray}
writing the wave function in polar form and assuming a Gaussian ansatz for the probability density 
\begin{equation} \label{eq: Gauss_an}
 |\psi_{v_0, \ga}(x, t)|^2 = \frac{1}{\sqrt{2\pi} \si_{\ga}(t)} \exp \left[ -\frac{(x-x_{\ga}(t))^2}{2 \si_{\ga}(t)^2} \right] ,
\end{equation}
the center of the wave packet and its width are expressed as
\begin{eqnarray} 
x_{\ga}(t) &=& - \frac{ K }{ m \om^2 } + \left( x_0 + \frac{ K }{ m \om^2 } \right) \left[ \cosh \Om t + \frac{\ga}{2} \frac{\sinh \Om t}{\Om} \right] e^{-\ga t /2} + v_0 ~ \frac{\sinh \Om t}{\Om}~e^{-\ga t /2} ,
\label{eq: xbar}
\\
\si_{\ga}(t) &=& \sigma_0~e^{-\ga t /2}~ 
\sqrt{ \left( \cosh \Om t + \frac{\ga}{2} \frac{\sinh \Om t}{\Om} \right)^2 + \frac{\hb^2}{4m^2 \sigma_0^4} \frac{\sinh^2 \Om t}{\Om^2} } \label{eq: CK_width_quandratic}
\end{eqnarray}
where the frequency $\Om$ is defined by
\begin{eqnarray} \label{eq: Omega}
\Om &=& \sqrt{\om^2 + \ga^2/4} .
\end{eqnarray}
%

%\subsection{Schr\"odinger-Langevin equation}

On the other hand, the so-called Schr\"{o}dinger-Langevin or Kostin nonlinear (logarithmic) equation for the Ohmic case is written as  
\begin{eqnarray}  \label{eq: Sch_Lan}
i \hbar \frac{\partial}{\partial t}\psi_{v_0, \ga}(x, t) &=& \left[ -\frac{\hbar^2}{2m} \frac{\partial^2}{\partial x^2}
+ V(x) + \frac{\gamma \hbar}{2 i} \left( \ln \frac{\psi_{v_0, \ga}}{\psi_{v_0, \ga}^*} - \left \langle \ln \frac{\psi_{v_0, \ga}}
{\psi_{v_0, \ga}^*} \right \rangle \right) 
\right] \psi_{v_0, \ga}(x, t) 
\end{eqnarray}
and following the same procedure as before, $ x_{\ga}(t) $ has the same expression given by Eq. (\ref{eq: xbar}) while the width is the solution 
of the generalized Pinney equation \cite{Pinney,NaMi-book-2017}
\begin{eqnarray} \label{eq: Pinney}
\ddot{ \si }_{\ga}(t) + \ga \dot{ \si }_{\ga}(t) - \frac{\hb^2}{4 m^2  \si_{\ga}(t)^3} - \om^2  \si_{\ga}(t)  &=& 0  
\end{eqnarray}
which has no analytical solution.

Thus, the discrepancy between both approaches relies on the behavior of the width of the thermal distribution. As discussed previously
\cite{MoMi-AP-2018}, when the interaction with an environment is considered, linear quantum mechanics is no longer applicable. This linear
CK approach is seen more like an effective approach to dissipation. The Kostin approach comes from the standard Langevin
equation which is also issued from the Caldeira-Leggett Hamiltonian.  

\subsection{Effect of dissipation on the width of the thermal Gaussian wave packet}

As we have already mentioned, the diagonal elements of the density operator have the interpretation of probability distribution. 
In a dissipative medium, the time evolution of the state (\ref{eq: den_mat_0}) yields to 
\begin{eqnarray} \label{eq: prob_den_gamma}
\rho_{\ga, T}(x, t) &=& \la x | \rho_{\ga, T}(t) | x \ra
= \frac{ 1 }{ \sqrt{2\pi} \si_{\ga, T}(t) } 
\exp \left[ -\frac{( x- X_{\ga}(t) )^2}{2 \si_{\ga, T}(t)^2} \right] ,
\end{eqnarray}
for the thermal probability density, where
\begin{eqnarray} 
X_{\ga}(t) &=&  - \frac{ K }{ m \om^2 } + \left( x_0 + \frac{ K }{ m \om^2 } \right) \left[ \cosh \Om t + \frac{\ga}{2} \frac{\sinh \Om t}{\Om} \right] e^{-\ga t /2}  \label{eq: Xt_gamma}
\\
\si_{\ga, T}(t) &=& \sqrt{ \si_{\ga}(t)^2 + e^{-\ga t} ~ \frac{k_B T}{m \Om^2} \sinh^2(\Om t)} \label{eq: var}     .
\end{eqnarray}

It is clear from the previous equations that the temperature-dependence of the wave packet is independent of the approach we use 
for taking into account the dissipation.
For the free case, $ \Om = \ga / 2 $ and the explicit form of the variance of the thermal wave packet in the CK framework reads as
\begin{eqnarray} \label{eq: var_free_CK}
\si_{\ga, T}(t)^2 &=& \si_0^2 + \frac{ \hb^2 } { 4m^2 \si_0^2 \ga^2 } ( 1 - e^{-\ga t} )^2 
+ e^{-\ga t} ~ \frac{4 k_B T}{m \ga^2} \sinh^2 \left( \frac{\ga t}{2} \right)   .
\end{eqnarray}
In the study of decoherence, Ford and O'Connell \cite{FoCo-JOB-2003} obtained 
\begin{eqnarray} \label{eq: Ford_var}
w^2(t) &=& \sigma_0^2 - \frac{ [\hat{x}(0), \hat{x}(t)]^2 }{4\sigma_0^2} + \langle  ( \hat{x}(t) - \hat{x}(0) )^2  \rangle
\end{eqnarray}
for the variance of the the probability distribution at time $t$, taking the initial state as a Gaussian wave packet. The last term which is the 
mean square displacement is temperature-dependent, while the second term is not. One has that 
$ [\hat{x}(0), \hat{x}(t)] = i \hbar ( 1 - e^{-\ga t} ) / m \ga $
and $ \langle  ( \hat{x}(t) - \hat{x}(0) )^2  \rangle = \frac{2 k_B T}{m \ga} \left( t - \frac{1 - e^{-\ga t}}{\ga} \right) $ for high temperatures, $ k_B T \gg \hb \ga $. 
Apart from different physical contexts, the comparison of Eqs. (\ref{eq: var_free_CK}) and (\ref{eq: Ford_var}) shows that the 
first two terms of them are exactly the same.

%===========================
\section{Transmission through a parabolic repeller} \label{sec: tran_rep}
%===========================

\subsection{Thermal transmission probability with dissipation}

Now consider transmission of our ensemble of particles through the parabolic repeller $ V(x) = - \frac{1}{2} m \om^2 x^2 $. The 
transmission probability for each element of our ensemble $\psi_{v_0, \ga}(x, t)$ is given by \cite{Pa-JPA-1997, BaJa-JPA-1992, Pa-JPA-1990} 
\begin{eqnarray} \label{eq: tran_prob_psi}
P_{\tr}(t; v_0, \ga) &=& \frac{ \erf( x_{\ga}(t) / \sqrt{2} \si_{\ga}(t)) - \erf(x_0 / \sqrt{2} \si_0) }{ \erfc(x_0 / \sqrt{2} \si_0)  }
\end{eqnarray}
where $ x_{\ga}(t) $ is given by Eq. (\ref{eq: xbar}) with $ K = 0 $ and $ \si_{\ga}(t) $ by Eq. (\ref{eq: CK_width_quandratic}) in 
the CK framework or by the solution of the generalized Pinney equation (\ref{eq: Pinney}) in the Kostin one.
It should be noted that in a transmission process, the incoming wave packet is initially well-localized on the {\it left}, $ x_0 < 0 $, of the 
barrier. In such a case,  $\si_0 \ll |x_0|$, one has
\begin{eqnarray*}
\erf(x_0 / \sqrt{2} \si_0) \approx -1, \qquad \erfc(x_0 / \sqrt{2} \si_0) \approx 2
\end{eqnarray*}
from which  Eq. (\ref{eq: tran_prob_psi}) can be rewritten as
\begin{eqnarray} \label{eq: tran_prob_psi_approx}
P_{\tr}(t; v_0, \ga) &\approx & \frac{1}{2}  \erfc \left( -\frac{x_{\ga}(t)}{\sqrt{2} \si_{\ga}(t)} \right)   .
\end{eqnarray}
Now, due to the Maxwell-Boltzmann distribution function (\ref{eq: Maxwell_Boltz_dis}) for the initial velocities, the time dependent 
thermal transmission probability under the presence of dissipation is given by
\begin{eqnarray} \label{eq: thermal_tran_prob}
P_{\tr}(t; \ga, T) &=& \int_{-\infty}^{\infty} dv_0 ~ f_T(v_0) ~ P_{\tr}(t; v_0, \ga)
\end{eqnarray}
and from the integral representation of the complementary error function \cite{NgGe-JRN-1969}
\begin{eqnarray} \label{eq: erfc_int_rep}
\erfc(z) &=& \frac{2}{\sqrt{\pi}} e^{-z^2} \int_0^{\infty} dy ~ e^{-(y^2+2zy)}  ,
\end{eqnarray}
one can express the corresponding transmission probability as
\begin{eqnarray} 
P_{\tr}(t; \ga, T) &=& \frac{1}{\pi} \sqrt{ \frac{m}{2 k_B T} } \int_0^{\infty} dy ~ e^{-y^2}
\int_{-\infty}^{\infty} dv_0 ~ \exp \left[ - \frac{mv_0^2}{2k_B T} - \frac{x_{\ga}(t)^2}{2 \si_{\ga}(t)^2 } + \sqrt{2} \frac{x_{\ga}(t)}{\si_{\ga}(t)} y \right] \nonumber \\
&=& \frac{1}{2} \erfc \left( - \frac{ X_{\ga}(t) }{\sqrt{2} \si_{\ga, T}(t)} \right) 
\label{eq: thermal_tran_prob_1}
\end{eqnarray}
where $ X_{\ga}(t) $ is given by Eq. (\ref{eq: Xt_gamma}) with $ K = 0 $ and $ \si_{\ga, T}(t) $ by Eq. (\ref{eq: var}).

An alternative way to compute the thermal transmission probability is to make use of $\rho_T(x, t)$.  
Thus, the thermal transmission probability should be given now by
\begin{eqnarray} \label{eq: thermal_tran_prob2}
P_{\tr}(t; \ga, T) &=& \int_0^{\infty} dx ~ \rho_{\ga, T}(x, t) = 
\frac{1}{2} \erfc \left( - \frac{ X_{\ga}(t) }{\sqrt{2} \si_{\ga, T}(t)} \right)
\end{eqnarray}
as should be. The stationary value of the thermal transmission probability is reached when 
\begin{eqnarray} \label{eq: thermal_tran_prob_st}
P_{\tr}(\ga, T) &=& P_{\tr}(t; \ga, T) \bigg|_{t\rightarrow \infty} 
\end{eqnarray}
which in the CK approach reduces to 
\begin{eqnarray} \label{eq: thermal_tran_prob_st_CK}
P_{\tr}(\ga, T) \bigg|_{\CK} &=& 
\frac{1}{2} \erfc \left( \frac{ - x_0 ( 1 + \frac{\ga}{2\Om} ) }
{\sqrt{2} \si_0 \sqrt{
( 1 + \frac{\ga}{2\Om} )^2+  \frac{ \hb^2 }{ 4 m^2 \si_0^4 \Om^2 } + \frac{k_B T}{m \si_0^2 \Om^2 } } }   .
\right)
\end{eqnarray}
%

%===========================
\subsection{Thermal characteristic times with dissipation}
%===========================

%As we aim to study characteristic times in the framework of Bohmian mechanics, we briefly review main concepts of this view.
%
In the context of Bohmian mechanics, the complete description of a system is given by its wave function and its position in configuration 
space. As usual, the evolution of the wave function is given by the Schr\"{o}dinger equation but particle trajectories are specified 
through the so-called guidance equation
\begin{eqnarray} \label{eq: guidance}
\frac{dx}{dt} &=& \frac{\hb}{m} ~ \text{Im} \left\{ \frac{ \pa_x \psi(x, t) }{\psi(x, t)} \right\} \bigg|_{x = x(t)} \qquad 
\end{eqnarray}
where $ \pa_x = \pa / \pa x $ and $ x(t) $ is the Bohmian or quantum trajectory. To be clear, in the following, Bohmian trajectories 
will be labelled by $ x(x^{(0)}, t; v_0, \ga) $ when considering a Gaussian ansatz where the center of the corresponding wave packet moves 
with the initial velocity $ v_0 $ in a viscous medium with friction $\ga$ and $ x^{(0)} $ is the initial position of the Bohmian particle. The 
general expression for this Bohmian trajectory assuming the Gaussian ansatz and for potentials up to second order is written as
\cite{NaMi-book-2017}
\begin{eqnarray} \label{eq: btray}
x(x^{(0)}, t; v_0, \ga) &=& x_{\ga}(t)  + (x^{(0)}- x_0) \frac{\si_{\ga}(t)}{\si_{\ga}(0)} .
\end{eqnarray}
%

%===========================
\subsubsection{Arrival times}
%===========================

%
In this context, it was proved by Leavens \cite{Le-book-2002} from the non-crossing property of Bohmian trajectories that the arrival 
time distribution for those particles that actually reach the detector is proportional to the modulus of the probability current density. 
For  $ \psi_{v_0, \ga}(x, t) $, the arrival time distribution at the detector location $ x_{\xd} $ is given by
\begin{eqnarray} \label{eq: pcd_psi}
\Pi_{\text{A}}(x_{\xd} , t; v_0, \ga) &=& \frac{ | j_{v_0, \ga}(x_{\xd} , t) | }{ \int_0^{\infty} dt' ~ | j_{v_0, \ga}(x_{\xd} , t') | }  
\end{eqnarray}
and therefore for a pure ensemble described by the wave function $ \psi_{v_0, \ga}(x, t) $, the mean arrival time at the detector location 
can be written as 
\begin{eqnarray} \label{eq: mean_ar_psi}
\uptau_{\text{A}}(x_{\xd}; v_0, \ga) &=&  \int_0^{\infty} dt' ~ t' ~ \Pi_{\text{A}}(x_{\xd} , t; v_0, \ga) .  
\end{eqnarray}
By averaging now over the Maxwell-Boltzamnn distribution, one can calculate thermal mean arrival times when the system is described by 
Eq. (\ref{eq: mat_elem_t}) as
\begin{eqnarray} \label{eq: mean_ar_thermal}
\uptau_{\text{A}}(x_{\xd}; \ga, T) &=&  \int dv_0 ~ f_T(v_0) ~ \uptau_{\text{A}}(x_{\xd}; v_0, \ga) 
\\
&=& \int_0^{\infty} dt' ~ t' \int dv_0 ~ f_T(v_0)~ \Pi_{\text{A}}(x_{\xd} , t'; v_0, \ga) 
\end{eqnarray}
from which one obtains
\begin{eqnarray} \label{eq: pcd_thermal}
\Pi_{\text{A}}(x_{\xd} , t; \ga, T) &=& 
\frac{ \int dv_0 ~ f_T(v_0)~ \Pi_{\text{A}}(x_{\xd} , t; v_0, \ga) }
{ \int_0^{\infty} dt' \int dv_0 ~ f_T(v_0)~ \Pi_{\text{A}}(x_{\xd} , t'; v_0, \ga) }.
\end{eqnarray}
for the thermal arrival time distribution with dissipation.

%

%
%We have shown in appendix \ref{app: den_mat} that the thermal PCD
%
%\begin{eqnarray} \label{eq: pcd_thermal}
%j_{\ga, T}(x, t) &=& \int dv_0 f_T(v_0) ~ j_{v_0, \ga}(x, t)
%\end{eqnarray}
%
%satisfies the continuity equation
%
%\begin{eqnarray} \label{eq: coneq_thermal}
%\frac{\pa}{\pa x} \rho_{\ga, T}(x, t) + \frac{\pa}{\pa x}j_{\ga, T}(x, t) &=& 0 .
%\end{eqnarray}
%

%Therefore, mean arrival time at the detector location $x_{\xd}$ is given by
%
%\begin{eqnarray} \label{eq: mean_ar_thermal}
%\uptau_{\text{A}}(x_{\xd}; \ga, T) &=&  \int_0^{\infty} dt' ~ t' ~ \Pi_{\text{A}}(x_{\xd}, t'; \ga, T) 
%\end{eqnarray}
%
%where 
%
%\begin{eqnarray} \label{eq: ar_dis_thermal}
%\Pi_{\text{A}}(x_{\xd}, t'; \ga, T) &=& \frac{ | j_{\ga, T}(x_{\xd}, t) | }
%{ \int_0^{\infty} dt' ~ | j_{\ga, T}(x_{\xd}, t') |  }
%\end{eqnarray}
%
%is the arrival time distribution. 

\subsubsection{Thermal dwelling, transmission and reflection times}

%We now study a scattering problem in which an ensemble of particles described by the wave function $\psi_{v_0, \ga}(x, 0)$ is incident from the left on a potential barrier. 
In the Bohm trajectory context, characteristic times are also important issues \cite{Le-LNP-2008}. Here, we would like to generalize this 
discussion when dealing with a mixed ensemble which is described by the the density matrix, Eq. (\ref{eq: den_mat_t}). 
The time that a particle, with initial position $ x^{(0)} $, spends in a given space interval $[x_1, x_2]$ can be expressed as
\begin{eqnarray}
t(x_1, x_2; x^{(0)}, v_0, \ga) &=& \int_0^{\infty} dt~ \theta( x(x^{(0)}, t; v_0, \ga) - x_1 ) ~ \theta( x_2 -x(x^{(0)}, t; v_0, \ga) )
\\
&=& \int_0^{\infty} dt \int_{x_1}^{x_2} dx ~ \delta( x(x^{(0)}, t; v_0, \ga) - x ) 
\end{eqnarray}
where the $\theta(x)$ is the step function. Then, the mean dwelling time is readily calculated as 
\begin{eqnarray} \label{eq: dwell_time}
\uptau_D(x_1, x_2; v_0, \ga) &=& \int_{-\infty}^{\infty} dx^{(0)}~ |\psi_{v_0, \ga}(x^{(0)}, 0)|^2 t(x_1, x_2; x^{(0)}, v_0, \ga)  
= \int_0^{\infty} dt \int_{x_1}^{x_2} dx ~ |\psi_{v_0, \ga}(x, t)|^2 ,
\end{eqnarray}
where, in the second equality, we have used the fact that $ |\psi_{v_0, \ga}(x, t)|^2 = \int dx^{(0)}~ |\psi_{v_0, \ga}(x^{(0)}, 0)|^2 
\delta( x(x^{(0)}, t; v_0, \ga) - x )$. 
For a one-dimensional motion, due to the non-crossing property of Bohmian trajectories, there is always a critical trajectory 
$ x_c(t; v_0, \ga) $ which separates transmitted from reflected trajectories in a scattering problem \cite{SaMi-AOP-2013}. 
Thus, for the {\it stationary} transmission probability one can always write
\begin{eqnarray} \label{eq: tr_prob_BM}
P_{\tr}(v_0, \ga) &=& \int_{ x_c(t; v_0, \ga) }^{\infty} dx~|\psi_{v_0, \ga}(x, t)|^2 .
\end{eqnarray}
By introducing now
\begin{eqnarray} 
1 &=& \theta( x - x_c(t; v_0, \ga)) + \theta( x_c(t; v_0, \ga) - x ) 
\end{eqnarray}
in Eq. (\ref{eq: dwell_time}), the dwelling time can be split as  
\begin{eqnarray} \label{eq: dwell_time_BM1}
\uptau_D(x_1, x_2; v_0, \ga) &=& P_{\tr}(v_0, \ga) ~ \uptau_{\tr}(x_1, x_2; v_0, \ga) + P_{\re}(v_0, \ga) ~\uptau_{\re}(x_1, x_2; v_0, \ga)  
\end{eqnarray}
where the transmission and reflection times are defined respectively as follows
\begin{eqnarray} 
\uptau_{\tr}(x_1, x_2; v_0, \ga) &=& \frac{1}{P_{\tr}(v_0, \ga)} \int_0^{\infty} dt \int_{x_1}^{x_2} dx ~ 
|\psi_{v_0, \ga}(x, t)|^2 ~ \theta( x - x_c(t; v_0, \ga)) \label{eq: tr_time_BM1} 
\\
\uptau_{\re}(x_1, x_2; v_0, \ga) &=& \frac{1}{P_{\re}(v_0, \ga)} \int_0^{\infty} dt \int_{x_1}^{x_2} dx ~ 
|\psi_{v_0, \ga}(x, t)|^2 ~ \theta( x_c(t; v_0, \ga) - x ) \label{eq: ref_time_BM1}
\end{eqnarray}
in terms of the reflection and transmission probabilities with  $ P_{\re}(v_0, \ga) = 1 - P_{\tr}(v_0, \ga) $.
These relations show that the calculation of characteristic times within the Bohmian mechanics just requires the knowledge of 
the critical trajectory $ x_c(t; v_0, \ga) $. However, it has been proved that there is no need to compute this single trajectory. 
One can always write  \cite{Kr-JPA-2005}
\begin{eqnarray} 
\uptau_{\text{D}}(x_1, x_2; v_0, \ga) &=& \int_0^{\infty} dt [Q(x_1, t; v_0, \ga) - Q(x_2, t; v_0, \ga)] \label{eq: dwell_time_BM}
\\
\uptau_{\tr}(x_1, x_2; v_0, \ga) &=& \frac{1}{P_{\tr}(v_0, \ga)} \int_0^{\infty} dt 
~\left[ \text{min}\{ Q(x_1, t; v_0, \ga), P_{\tr}(v_0, \ga) \} - \text{min}\{ Q(x_2, t; v_0, \ga), P_{\tr, \ga}(v_0, \ga) \} \right]
 \label{eq: tr_time_BM} 
\\
\uptau_{\re}(x_1, x_2; v_0, \ga) &=& \frac{1}{P_{\re}(v_0, \ga)} \int_0^{\infty} dt 
~\left[ \text{max}\{ Q(x_1, t; v_0, \ga), P_{\tr}(v_0, \ga) \} - \text{max}\{ Q(x_2, t; v_0, \ga), P_{\tr}(v_0, \ga) \} \right]
 \label{eq: ref_time_BM}
\end{eqnarray}
where
\begin{eqnarray} 
Q(x, t; v_0, \ga) &=& \int_x^{\infty} dx'~|\psi_{v_0, \ga}(x', t)|^2 = \int_0^t dt'~ j_{v_0, \ga}(x, t') 
\label{eq: Q_v0_BM}
\\
&=& \frac{1}{2} \erfc \left( \frac{x - x_{\ga}(t)}{ \sqrt{2} \sigma_{\ga}(t) }  \right)
\label{eq: Q_v0_BM2}
\end{eqnarray}
is the probability of being the particle beyond the point $x$. %In the Bohmian framework term ``finding" must be replaced by ``being".
For a mixed ensemble of non-interacting particles, which is initially described by the density operator (\ref{eq: den_mat_t}), the thermal averaging of
Eq. (\ref{eq: dwell_time}) over the Maxwell-Boltzmann distribution  yields
\begin{eqnarray} \label{eq: tauD_T1}
\uptau_D(x_1, x_2; \ga, T) &=& \int_0^{\infty} dt ~ P_{\ga, T}(x_1, x_2, t)
\end{eqnarray}
where
\begin{eqnarray} \label{eq: prob_x1x2}
P_{\ga, T}(x_1, x_2, t) &=& \int_{x_1}^{x_2} dx ~ \rho_{\ga, T}(x_1, x_2, t)
\end{eqnarray}
gives the probability for being the particle in the space interval $[x_1, x_2]$ at time $t$ and at the temperature $T$.

%Averaging (\ref{eq: dwell_time_BM1}) over the Maxwell-Boltzmann distribution (\ref{eq: Maxwell_Boltz_dis}) leads to
%
%\begin{eqnarray}
%\uptau_D(x_1, x_2; \ga, T) &=& \int_{-\infty}^{\infty} dv_0 ~ f_T(v_0) P_{\tr}(v_0, \ga) ~ \uptau_{\tr}(x_1, x_2; v_0, \ga) + \int_{-\infty}^{\infty} dv_0 ~ f_T(v_0)  P_{\re}(v_0, \ga) ~\uptau_{\re}(x_1, x_2; v_0, \ga) 
%\\
%& \equiv &
%P_{\tr}(\ga, T) ~ \uptau_{\tr}(x_1, x_2; \ga, T) + P_{\re}(\ga, T) ~\uptau_{\re}(x_1, x_2; \ga, T)
%\end{eqnarray}
%
%where we have defined thermal transmission and reflection times as
%
%\begin{eqnarray*}
%\uptau_{\tr}(x_1, x_2; \ga, T) &=& \frac{1}{P_{\tr}(\ga, T)} \int_{-\infty}^{\infty} dv_0 ~ f_T(v_0) \int_0^{\infty} dt \left[ \text{min}\{ Q(x_1, t; v_0, \ga), P_{\tr}(v_0, \ga) \} - \text{min}\{ Q(x_2, t; v_0, \ga), P_{\tr}(v_0, \ga) \} \right]
%\\
%\uptau_{\re}(x_1, x_2; \ga, T) &=& \frac{1}{P_{\re}(\ga, T)} \int_{-\infty}^{\infty} dv_0 ~ f_T(v_0) \int_0^{\infty} dt \left[ \text{max}\{ Q(x_1, t; v_0, \ga), P_{\tr}(v_0, \ga) \} - \text{max}\{ Q(x_2, t; v_0, \ga), P_{\tr}(v_0, \ga) \} \right]
%\end{eqnarray*}
%

By averaging Eq. (\ref{eq: dwell_time_BM}), one has that
\begin{eqnarray} \label{eq: tauD_T2}
\uptau_D(x_1, x_2; \ga, T) &=& \int_0^{\infty} dt ~ [ Q_{\ga, T}(x_1, t) - Q_{\ga, T}(x_2, t) ]
\end{eqnarray}
an alterntive expression for the the dwelling time with 
\begin{eqnarray} \label{eq: Q_thermal_def}
Q_{\ga, T}(x, t) &=& \int_{-\infty}^{\infty} dv_0 ~ f_T(v_0) ~ Q(x, t; v_0, \ga) 
\end{eqnarray}
Again, by using Eq. (\ref{eq: erfc_int_rep}) for the integral representation of the complementary error function, 
Eq. (\ref{eq: Q_thermal_def}) can be rewritten as
\begin{eqnarray} \label{eq: Q_thermal}
Q_{\ga, T}(x, t) &=& \frac{1}{2} \erfc \left( \frac{ x - x_0 \cosh(\om t) }{\sqrt{2} ~ \si_{\ga, T}(t)} \right) .
\end{eqnarray}
One can easily check that the long-time limit of $ Q_{\ga, T}(x, t) $ for a given $x$ is just the stationary value for the thermal transmission probability 
(\ref{eq: thermal_tran_prob_st}).

The thermal transmission and reflection times in presence of dissipation are respectively given by
\begin{eqnarray}
\uptau_{\tr}(x_1, x_2; \ga, T) &=& \int dv_0~ f_T(v_0)~ \uptau_{\tr}(x_1, x_2; v_0, \ga)  \label{eq: tau_tr_T}
\\
\uptau_{\re}(x_1, x_2; \ga, T) &=& \int dv_0~ f_T(v_0)~ \uptau_{\re}(x_1, x_2; v_0, \ga) \label{eq: tau_ref_T}  .
\end{eqnarray}
It should be noted that at zero temperature, the Maxwell-Boltzmann distribution is just the Dirac delta function centered at 
$ v_0 = 0 $, 
\begin{eqnarray} \label{eq: MB_T=0}
f_T(v_0) \bigg|_{T=0} &=& \delta( v_0 )
\end{eqnarray}
meaning that instead of a mixed ensemble we have a pure ensemble where all elements of the ensemble are described by the same 
wave function $ \psi_{v_0 = 0, \ga}(x, t) $. In this case, thermal quantities are equivalent to those obtained for the pure ensemble with $ v_0 = 0 $.

%Clearly, an alternative way to define thermal transmission and reflection times is the following. As the thermal wave packet 
%$ \rho_{\ga, T}(x, t) $ has the interpretation of probability distribution, the thermal dwelling time (\ref{eq: tauD_T1}) can be decomposed
%as
%
%\begin{eqnarray} \label{eq: tauD_thermal_decom}
%\uptau_D(x_1, x_2; \ga, T) &=&
% P_{\tr}(\ga , T) \uptau'_{\tr}(x_1, x_2; \ga, T) + P_{\re}(\ga , T) \uptau'_{\re}(x_1, x_2; \ga, T)
%\end{eqnarray}
%
%where the new thermal transmission and reflection times are defined as 
%
%\begin{eqnarray*}
%\uptau'_{\tr}(x_1, x_2; \ga, T) &=& \frac{1}{P_{\tr}(\ga, T)} 
% \int_0^{\infty} dt \left[ \text{min}\{ Q_{\ga, T}(x, t), P_{\tr}(\ga, T) \} - \text{min}\{ Q_{\ga, T}(x, t), P_{\tr}(\ga, T) \} \right]
%\\
%\uptau'_{\re}(x_1, x_2; \ga, T) &=&  \frac{1}{P_{\re}(\ga, T)} 
%\int_0^{\infty} dt \left[ \text{max}\{ Q_{\ga, T}(x, t), P_{\tr}(\ga, T) \} - \text{max}\{ Q_{\ga, T}(x, t), P_{\tr}(\ga, T) \} \right]
%\end{eqnarray*}
%
%with $ P_{\re}(\ga, T) = 1 - P_{\tr}(\ga, T) $ being the thermal reflection probability.
%{\color{blue} One should note that there is nothing to do with trajectories in these definitions.}

%===============================================
\section{Results and discussion} \label{sec: nu_comp}
%===============================================

In order to simpify our calculations, we are going to work on dimensionless quantities. Thus, we use the following reference values:
$ \ti{t} = \dfrac{2 m \si_0^2}{\hb} $, $ \ti{\om} = \dfrac{1}{\ti{t}}$ and $ \ti{T} =  \dfrac{ \hb^2 }{ 4m\si_0^2 k_B }$ for times, frequencies
and temperatures, respectively. Then, we have that  $\bar{\ga}= \dfrac{\ga}{\ti{\om}} $,  $ \bar{\Om} = \dfrac{\Om}{\ti{\om}} $,  
$ \bar{T} = \dfrac{T}{\ti{T}} $. Moreover, lengths are also dimensionless when dividing by $ \si_0 $ and denoted by a bar symbol. 
In this way, Eq. (\ref{eq: thermal_tran_prob_st_CK}) for the stationary transmission probability for the CK approach takes the simple form
\begin{eqnarray} \label{eq: Tr_st_dimless}
P_{\tr}(\bar{\ga}, \bar{T}) &=&  \frac{1}{2} \erfc \left( \frac{ - \bar{x}_0 \left( 1 + \dfrac{\bar{\ga}}{2 \bar{\Om}} \right) }{\sqrt{2} \sqrt{ \left( 1 + \dfrac{\bar{\ga}}{2 \bar{\Om}} \right)^2 +  \dfrac{ 1 + \bar{T} }{ \bar{\Om}^2 } }}
\right)
\end{eqnarray}
where $ \bar{x}_0 = x_0 / \si_0  $. After the behavior of the complementary error function, Eq. (\ref{eq: Tr_st_dimless}) shows that 
the thermal transmission probability increases with temperature for a given friction $\ga$ and barrier's strength $\om$ and finally takes a
stationay value of $0.5$.
By taking the partial derivative of the argument of the complementary error function with respect to the barrier's strength and friction and noting 
the negative value of $x_0$, it is seen that the argument is an increasing function of $ \om $ (for a given temperature and friction) and also of $ \ga $ 
(for a given temperature and barrier's strength). Thus, the transmission probability also  decreases with both $ \om $ and $ \ga $. These results
are understandable because when one increases $ \om $, the parabolic barrier becomes more repulsive; whereas, when the friction increases, 
the interaction between particles and the environment also increases leading to more energy dissipation. 

For numerical calculations,  we use the mass of electron and the width of the initial wave packet to be $ \si_0 = 0.4 ~$\AA. 
Other parameters  chosen are: $ \bar{x}_0 = -20 $ for the center of the wave packet, $ \bar{\om} = 0.05 $ and $ \bar{\om} = 0.1 $ for the 
strengths of the barrier,  and $\bar{x}_{\xd} = 20$ for the detector location when computating the arrival times. 
For computing the thermal characteristic times, the interval $ [\bar{x}_1=-1, \bar{x}_2=1] $ is chosen.
To obtain thermal quantities one should integrate over all initial velocities with the Maxwell-Boltzmann distribution for a given temperature. 
In principle, any velocity should be included, even large negative values. However, due to the decomposition of the dwelling time into transmission 
and reflection times,  Eq. (\ref{eq: dwell_time_BM1}), this makes sense only when the transmission probability is not negligible. Thus, 
from a numerical point of view, the lower limit in the integration equals to a velocity for which this transmission probability is greater than or equal to 
$ 0.01 $. 
For the frictionless case, this requirement leads to $ \bar{v}_{0, \text{min}} \approx - 1.304  $ for $ \bar{\om} = 0.05 $ and 
$ \bar{v}_{0, \text{min}} \approx - 0.3111  $ for $ \bar{\om} = 0.1 $, where $ \bar{v}_0 = \frac{ v_0 }{ \ti{v}_0 } $ 
with $ \ti{v}_0 = \frac{\si_0}{\ti{t}} $ is the dimensionless velocity.

%===============================================
%Figure 
\begin{figure} 
\centering
\includegraphics[width=10cm,angle=-90]{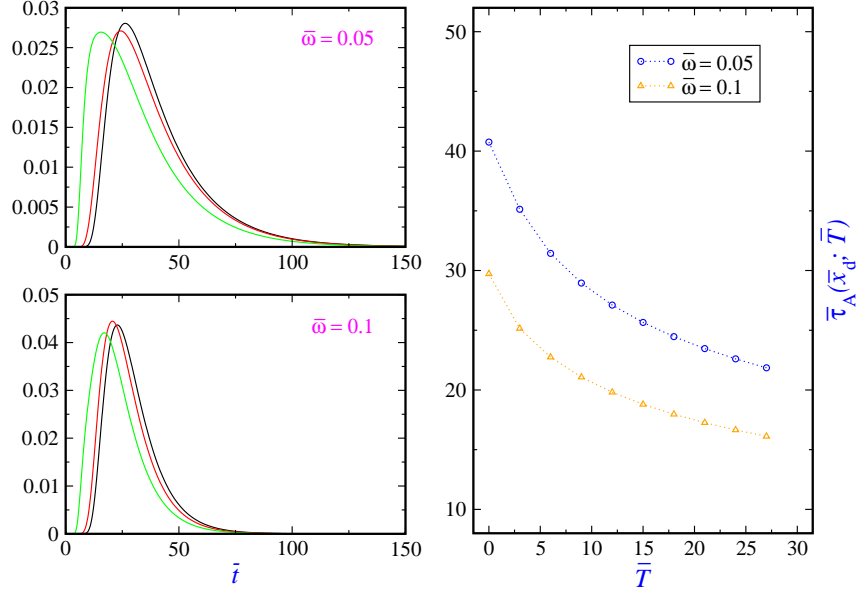}
\caption{(Color online) 
Arrival time distributions  for $\bar{ \om } = 0.05$ (left top panel) and $\bar{ \om } = 0.1$ (left bottom panel) with $ \ga = 0 $ and  
different values of temperature: $ \bar{T} = 0 $ (black curve), $ \bar{T} =  1 $ (red curve) and $ \bar{T} = 5 $ (green curve). 
Right panel displays mean arrival times at the detector location $\bar{x}_d = 20$ as a function of the temperature for two different parabolic barriers.
}
\label{fig: arrival} 
\end{figure}
%===============================================
%

In Figure \ref{fig: arrival}, arrival time distributions for $\bar{ \om } = 0.05$ (left top panel) and $\bar{ \om } = 0.1$ (left bottom panel) with 
$ \ga = 0 $ and  different values of temperature: $ \bar{T} = 0 $ (black curve), $ \bar{T} =  1 $ (red curve) and $ \bar{T} = 5 $ (green curve)
are plotted. In the right panel of the same figure, it is also displayed mean arrival times at the detector location as a function of the 
temperature for the two values of $\bar{\om}$. The maximum of the arrival time distribution moves to shorter times as temperature increases. 
For  a given temperature, this distribution becomes narrower with $ \bar{\om} $. 
As an expected result, the mean arrival time decreases with temperature and the strength of the barrier in this frictionless case. 

%===============================================
%Figure 
\begin{figure} 
	\centering
	\includegraphics[width=10cm,angle=-90]{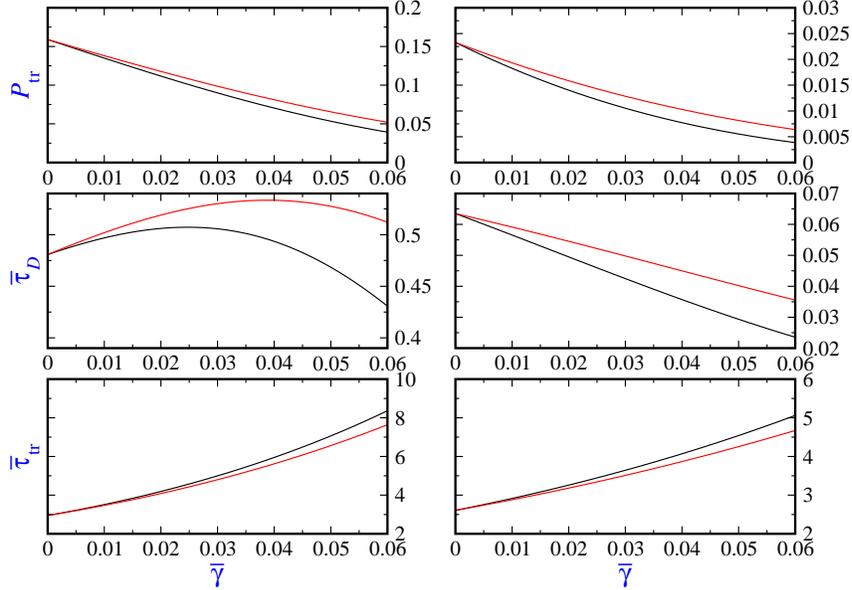}
	\caption{(Color online) 
		Transmission probability (top row), dwelling time (middle row) and transmission time (bottom row) versus friction for $ \bar{\om} = 0.05 $ 
		(left column) and $ \bar{\om} = 0.1 $ (right column) at zero temperature in the CK (black curves) and Kostin (red curves) approaches.
	}
	\label{fig: chartimes_T=0_gama}
\end{figure}
%===============================================

\begin{figure} 
\centering
\includegraphics[width=10cm,angle=-90]{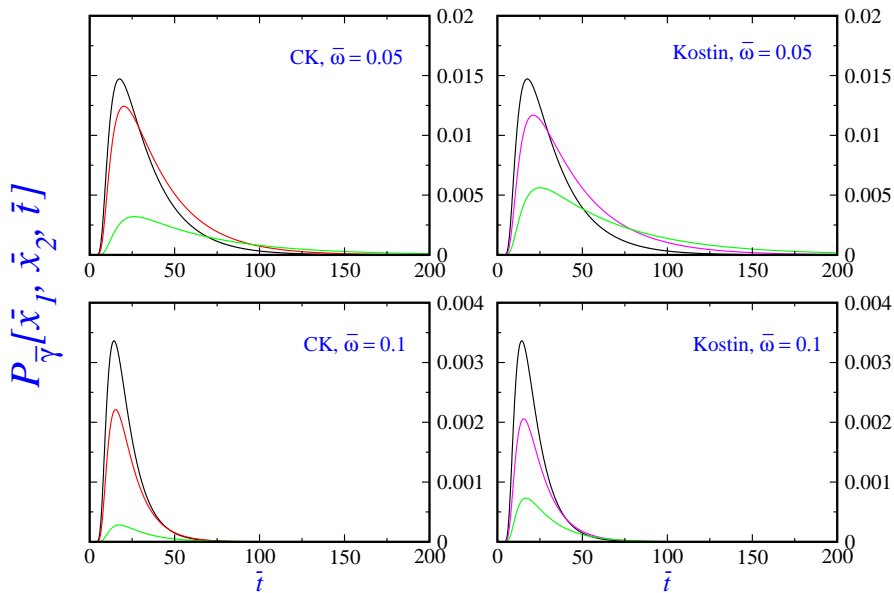}
\caption{(Color online) 
Probability of being the particle in the interval $[\bar{x}_1, \bar{x}_2]$ versus time for three different frictions $ \bar{\ga} = 0 $ (black curves), 
$ \bar{\ga} = 0.025 $ (red curves), $ \bar{\ga} = 0.04 $ (magenta curves) and $ \bar{\ga} = 0.1 $ (green curves) at zero temperature in the 
CK (left column) and Kostin (right column) approach and two parabolic barriers $ \bar{\om} = 0.05 $ (top row) and $ \bar{\om} = 0.1 $ 
(bottom row).
}
\label{fig: prob_T=0_gama}
\end{figure}

For comparison, the transmission probability (top row), dwelling time (middle row) and transmission time (bottom row) versus friction for 
$ \bar{\om} = 0.05 $ (left column) and $ \bar{\om} = 0.1 $ (right column) at zero temperature in the CK (black curves) and Kostin 
(red curves) approaches are displayed in Figure \ref{fig: chartimes_T=0_gama}. The discrepancies between both approaches are  rather important
although the global behavior is the same. As commented above, the Kostin values are more reliable than the CK ones. As is known, the transmission
probability decreases with friction. With $v_0=0$, the only contribution to the kinetic energy comes from the initial width of the Gaussian 
wave packet which is $\hbar^2 / 8 m \sigma_0^2$. 
For the values chosen for $\bar{\om}$, the expectation value of total energy is initially negative.
Thus, the dissipative dynamics develops only via {\it tunnelling}. Notice that this value is also important
because the transmission probabilities are one order of magnitude higher for $ \bar{\om} = 0.05 $. The dwelling time also changes dramatically
with the strength of the barrier, whereas the transmission time is smoother for both frequencies and of the same order.   
In figure \ref{fig: prob_T=0_gama}, the probability of being the particle in the interval $[\bar{x}_1, \bar{x}_2]$ versus time is shown 
for four different frictions $ \bar{\ga} = 0 $ (black curves), $ \bar{\ga} = 0.025 $ (red curves), $ \bar{\ga} = 0.04 $ (magenta curves) and 
$ \bar{\ga} = 0.1 $ (green curves) at zero temperature in the CK (left column) and Kostin (right column) approach and two parabolic barriers 
with frequencies $ \bar{\om} = 0.05 $ (top row) and $ \bar{\om} = 0.1 $ (bottom row). 
As this figure shows, fora given $ \ga $, probability increases with time, gets its maximum value in a time which depends on $ \ga $ 
and decreases afterwards. This dynamics describes the entrance of the Gaussian 
wave packet inside the interval $ [\bar{x}_1, \bar{x}_2] $ and then its leakage from this interval during the time. 
For $ \bar{\om} = 0.1 $ (bottom panels) curves with higher values of friction locate inside the curve for the frictionless one. Thus, the 
surface under the curves decreases with friction meaning dwelling time decreases with friction as the right middle panel of figure
 \ref{fig: chartimes_T=0_gama} shows. But, this is not true for $ \bar{\om} = 0.05 $.
In this case, dwelling time increases with friction at first, reaches its maximum value at $ \bar{\ga} \approx 0.025 $ for CK and $ \bar{\ga} \approx 0.04 $ for Kostin. 
% 
%===============================================
%Figure 
%===============================================

\begin{figure} 
\centering
\includegraphics[width=10cm,angle=-90]{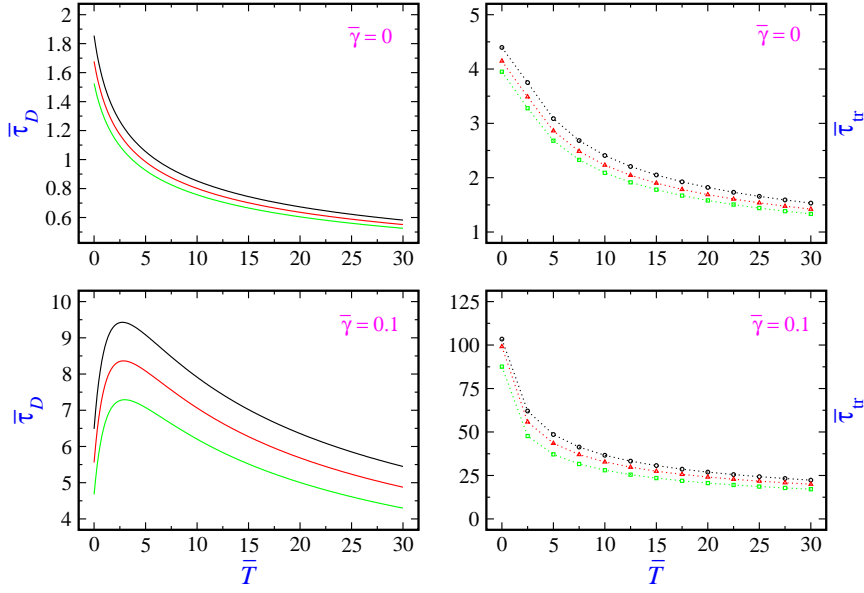}
\caption{(Color online) 
Thermal dwelling time (left column) and transmission time (right column) versus temperature for different values of parabolic repeller strength: 
$ \bar{\om} = 0.01 $ (black curves), $ \bar{\om} = 0.0125 $ (red curves) and $ \bar{\om} = 0.015 $ (green curves) and for $ \bar{\ga} = 0 $ 
(top row) and $ \bar{\ga} = 0.1 $ (bottom row) in the Kostin model.
}
\label{fig: tauD&tr_Ko} 
\end{figure}

Finally, in figure \ref{fig: tauD&tr_Ko} , the thermal dwelling (left column) and transmission (right column) times versus temperature 
are plotted for three different frequencies $\bar{\om}$ and two different frictions $\bar{\ga}$ for the Kostin approach.  This dissipative dynamics develops not only via tunnelling. As again expected,  both dwelling and transmission times decrease 
smoothly with $\bar{ \om} $ and temperature but increase with friction. The special case is for the bottom left panel where the thermal 
dwelling time displays a maximum at low temperatures for a friction of $0.1$; that is, the dwelling time is favoured at low temperatures. 
At these temperatures, the small velocities in both directions maintain the particle inside the barrier more time.   

%===============================================
%Figure 
%===============================================

In conclusion, along this work, we have presented and discussed thermal characteristic times for the dissipative dynamics under the 
presence of a parabolic repeller in terms of Bohmian trajectories. These thermal times as well as transmission probabilities have been 
defined and analyzed for several values of the frequency of the parabolic barrier and the CK and Kostin approaches within a linear and 
nonlinear framework, respectively. The thermal average in this work has been considered in a different way to that employed in the study 
of time-of-flight distributions for cold trapped atoms \cite{MaHoMaPa-PRA-2007} which for the probability current distribution is defined as 
\begin{eqnarray} \label{eq: pcd_thermal}
j_{\ga, T}(x, t) &=& \int dv_0 f_T(v_0) ~ j_{v_0, \ga}(x, t)
\end{eqnarray}
which is less convenient for a trajectory description. Similar results to previous works have been obtained. This work can be seen as the 
first step to deal with quantum stochastic dynamics within the Bohmian mechanics where the noise (thermal fluctuations) of the environment 
 is present. Work in this direction is now in progress.

\appendix

%========================================
\section{The continuity equation for a mixed ensemble} \label{app: den_mat}
%========================================

In the most general formulation of quantum systems, a quantum system is described by a density matrix
$\hat{\rho}$ instead of a state vector $ | \psi \rangle$. In this context, the von Neumann equation
\begin{eqnarray} \label{eq: voNewmann}
i \hb \frac{\pa \hat{\rho}}{\pa t} &=& [\hat{H}, \hat{\rho}]
\end{eqnarray}
has to be applied for the evolution of the system, where $H$ is the Hamiltonian of the system.

The expectation value of an observable $\hat{A}$ is computed as follows
\begin{eqnarray} \label{eq: obser_expect}
\langle \hat{A} \rangle(t)  &=& \text{Tr}( \hat{A} \hat{\rho}(t) )
\end{eqnarray}
where Tr means the trace operation. 
%For the space coordinate $\hat{x}$. From (\ref{eq: thermal_expect_value}) one has that
%
%\begin{eqnarray} \label{eq: x_expect_value}
%\langle \hat{x} \rangle(t) &=&  
%\int_{-\infty}^{\infty} dx ~ x ~ \langle x | \hat{\rho}(t) | x \rangle   .
%\end{eqnarray}
%
%This relation shows that the diagonal matrix elements of density operator has the interpretation of probability density.
In one dimension, the coordinate representation of Eq. (\ref{eq: voNewmann}) can be recast as
\begin{eqnarray} \label{eq: voNewmann_co_rep}
i \hb \frac{\pa}{\pa t} \rho(x, x', t) &=&  \left[ -\frac{\hb^2}{2m} \left( \frac{\pa^2}{\pa x^2} - \frac{\pa^2}{\pa x'^2}  \right) + V(x, t) - V(x',t) \right]  \rho(x, x', t)
\end{eqnarray}
with $ \rho(x, x', t) = \la x | \hat{\rho} | x' \ra $ and $V$ is the interaction potential.
From \cite{GoGrJo-CE-1991}, the current density matrix is
\begin{eqnarray} \label{eq: current_matrix}
j(x, x', t) &=& \frac{\hb}{m} \frac{1}{2i} \left( \frac{\pa}{\pa x} - \frac{\pa}{\pa x'} \right) \rho(x, x', t)
\end{eqnarray}
and the equation of motion (\ref{eq: voNewmann_co_rep}) can be expressed in the coordinate representation as 
\begin{eqnarray} \label{eq: evoleq2}
\frac{\pa \rho(x, x', t)}{\pa t} + \frac{\pa j(x, x', t)}{\pa x} + \frac{\pa j(x, x', t)}{\pa x'} + \frac{i}{\hb} (V(x, t)-V(x', t)) \rho(x, x', t) &=& 0 .
\end{eqnarray}
One notes that
\begin{eqnarray*} 
\frac{\pa j}{\pa x} + \frac{\pa j}{\pa x'} &=& 
\frac{\hb}{m} \frac{1}{2i} \left( \frac{\pa^2 \rho}{\pa x^2} - \frac{\pa^2 \rho}{\pa x'^2} \right)
= \frac{\hb}{m} \frac{1}{2i} \left( \frac{\pa^2 \rho}{\pa x^2} - \frac{\pa^2 \rho^*}{\pa x^2}\bigg|_{x' \leftrightarrow x} \right)
\end{eqnarray*}
where in the second equality we have used the fact that $ \rho(x, x', t) = \rho^*(x', x, t)$.
Thus, for $x'=x$ we have that
\begin{eqnarray*} 
\left( \frac{\pa j}{\pa x} + \frac{\pa j}{\pa x'} \right)\bigg|_{x'=x} &=& 
\frac{\pa}{\pa x} \text{Im} \left[ \frac{\hb}{m} \frac{\pa \rho}{\pa x}\bigg|_{x' = x} \right]
\end{eqnarray*}
Now, from (\ref{eq: evoleq2}) for $x'=x$, we have that
\begin{eqnarray*} 
\frac{\pa \rho(x, x', t)}{\pa t} \bigg|_{x'=x} + \frac{\pa}{\pa x} \text{Im} \left[ \frac{\hb}{m} \frac{\pa \rho}{\pa x}\bigg|_{x' = x} \right] &=& 0
\end{eqnarray*}
with
\begin{eqnarray}  
\rho(x, t) &=& \rho(x, x', t) \bigg|_{x'=x} \label{eq: rho} \\
j(x, t) &=& j(x, x', t) \bigg|_{x'=x}  \label{eq: j}
\end{eqnarray}
being the probability density function and probability current density, respectively. In the more familiar form, we have the continuity equation
\begin{eqnarray} \label{eq: coneq}
\frac{\pa \rho(x, t)}{\pa t} + \frac{\pa j(x, t)}{\pa x} &=& 0
\end{eqnarray}
%
%In the following we obtain continuity equation in another way by considering a mixed ensemble.
%

In a statistical mixture of states, the density operator is initially given by
\begin{eqnarray} \label{eq: mixed_state_0}
\hat{\rho}(0)  &=& \sum_i  w_i | \psi_i(0) \rangle \langle \psi_i(0) | , \qquad  \sum_i w_i = 1 .
\end{eqnarray}
where the coefficient $w_i$ gives the weight of the $ | \psi_i(0) \rangle $ state.
By using the evolution equation (\ref{eq: voNewmann}), the density matrix at time $t$ is then written as
\begin{eqnarray} \label{eq: mixed_state_t}
\hat{\rho}(t)  &=& \sum_i  w_i | \psi_i(t) \rangle \langle \psi_i(t) | ,
\end{eqnarray}
where $ | \psi_i(t) \rangle = e^{-i\hat{H}t/\hb} | \psi_i(0) \rangle  $. Now, due to the continuity equation for the component $ \psi_i(x, t)$
\begin{eqnarray} \label{eq: coneq_component_i}
\frac{\pa }{\pa t} |\psi_i(x, t)|^2 + \frac{\pa}{\pa x} \left( \frac{\hb}{m} \Ima
\left[ \psi_i^*(x, t) \frac{\pa}{\pa x} \psi_i(x, t) \right] \right) &=& 0   ,
\end{eqnarray}
one has that
\begin{eqnarray} \label{eq: coneq_mixed_rho}
\frac{\pa}{\pa t} \sum_i  w_i |\psi_i(x, t)|^2 + \frac{\pa}{\pa x} \left( \frac{\hb}{m} \Ima
\left[ \sum_i  w_i \psi_i^*(x, t) \frac{\pa}{\pa x} \psi_i(x, t) \right] \right) &=& 0 .
\end{eqnarray}
with
\begin{eqnarray} \label{eq: rho_mixed}
\rho(x, t)  &=& \sum_i  w_i |\psi_i(x, t)|^2 
\end{eqnarray}
and
\begin{eqnarray} \label{eq: cur_mixed}
j(x, t)  &=& \frac{\hb}{m}  \sum_i  w_i  ~ \Ima
\left[\psi_i^*(x, t) \frac{\pa}{\pa x} \psi_i(x, t) \right],
\end{eqnarray}
%
%which noting the relation $ \rho(x, x', t) = \sum_i  w_i \psi_i^*(x', t) \psi_i(x, t) $, coincide respectively with the general formulas 
%(\ref{eq: rho}) and (\ref{eq: j}). 
%=====================================================

\vspace{2cm}
\noindent
{\bf Acknowledgements}
\vspace{1cm}

SVM acknowledges support from the University of Qom and SMA support from 
the Ministerio de Ciencia, Innovaci\'on y Universidades (Spain) under the Project 
Project FIS2017-83473-C2-1-P.

%=====================================================

%===========================

%
\end{document}